\documentclass[preprint,12pt]{elsarticle}

\usepackage{amssymb}
\biboptions{numbers,sort&compress}
\usepackage[colorlinks, linkcolor=red, anchorcolor=blue, citecolor=green]{hyperref}

\usepackage[]{caption2}

\usepackage{multirow}
\usepackage{lineno}
\journal{Computational Materials Science}

\begin{document}

\begin{frontmatter}



\title{Effects of transmutation elements in tungsten}


\author[label1,label2]{Qiang Zhao\corref{cor1}}
\author[label1,label2]{Zheng Zhang}
\author[label1,label2]{Mei Huang}
\author[label2,label3,label4]{Xiaoping Ouyang}
\cortext[cor1]{Corresponding author. E-mail address: qzhao@ncepu.edu.cn (Qiang Zhao).}
\address[label1]{Beijing Key Laboratory of Passive Safety Technology for Nuclear Energy, North China Electric Power University, Beijing 102206, PR China}
\address[label2]{School of Nuclear Science and Engineering, North China Electric Power University, Beijing 102206, PR China}
\address[label3]{Northwest Institute of Nuclear Technology, Xi'an 710024, PR China}
\address[label4]{School of Materials Science and Engineering, Xiangtan University, Xiangtan 411105, PR China}
\begin{abstract}
Tungsten (W) is widely considered as the most promising plasma facing material (PFM), which will be used in nuclear fusion devices. Due to the transmutation reaction caused by the fusion neutron irradiation, transmutation elements (such as Re, Os, and Ta) are generated in the W-based PFM during the operation of nuclear fusion devices. In this paper, transmutation elements effects on mechanical properties of the W and the behavior of hydrogen/helium (H/He) atom in the W were investigated by using the first-principles calculation method. The results show that the ductility of the W is enhanced by transmutation elements if dislocation and other defects were ignored, while mechanical properties of the W incompletely depend on transmutation elements content. Compared with the pure W, the formation energy of the H/He in the W is reduced by transmutation elements, but the most favorable sites of the H/He in the W is not changed. Except for a repulsion between Ta and He in the W, the H/He in the W is attracted by transmutation elements. In addition, transmutation elements can change the best diffusion path of the H/He in the W and increase the diffusion rate of the H/He in W. This study provides a theoretical basis for the Tungsten (W) application, and further study on the effects of transmutation elements in the W will be needed.
\end{abstract}


\end{frontmatter}


\section{Introduction}
Nuclear fusion is considered as the most promising sustainable energy source, and the best solution for the future energy crisis in the world\cite{Lawson1957Power}. The selection of structural materials plays a decisive role in the success of the controlled nuclear fusion, especially the plasma facing material(PFM). Tungsten (W) is widely considered as the greatest potential PFM which will be used in the fusion reactor\cite{Philipps2011Tungsten}. Naturally, the W contains five stable isotopes: $^{180}\rm W(0.1\%)$, $^{182}\rm W(26.3\%)$, $^{183}\rm W(14.3\%)$, $^{184}\rm W(30.7\%)$, and $^{186}\rm W(28.6\%)$. Under the fusion neutron irradiation, rhenium (Re) and osmium (Os) isotopes are produced in the W because of the ($n$, $\gamma$) and ($n$, $2n$) transmutation reactions. Tantalum (Ta) is generated in the W through the $\beta^{+}$ decay of $^{181}\rm{W}$ after the $^{180}\rm W$ atom absorbs a neutron\cite{Cottrell2006Transmutation,Gilbert2011Neutron}. The generation of transmutation elements (TEs) eventually leads to the formation of the brittle intermetallic phase in the W\cite{Hasegawa2014Neutron,Hasegawa2015Neutron}, and this phenomenon is bad for the W to be used as the PFM. As part of TEs, Re, Os, and Ta are mainly studied in our research.

With the rapid development of computer technology, the first-principles calculation method has been widely used in material research\cite{freysoldt2014first,wimmer1995computational,li2018effects,li2017dehydrogenation}. Through this method, researches on TEs effects on the performance of the W have been conducted with a great success, such as the influence of the concentration of Re in W on thermodynamic properties of W-Re alloys\cite{chouhan2012ab}, the effects of Re and Ta on the ideal tensile strength of W\cite{Giusepponi2013The}, the structural stability and mechanical properties of W-Re and W-Ta alloys\cite{Wei2014First}, the crystal structure and mechanical properties of W-Re-Os system\cite{Li2016Ab}, the elastic properties of sigma phase W-Re alloy\cite{Bonny2017Elastic}, the interaction between the interstitial cluster and TEs in W\cite{Setyawan2016Ab}, the effect of stress on the diffusion of Re in W-Re alloy\cite{Hossain2014Stress}, and the abnormal gathered behavior of Re in W-Re alloy contains some vacancies\cite{wrobel2017first}. These researches are helpful for the development of the W-based PFM, contributing to TEs effects on mechanical properties of the W and the aggregation of the TEs in the W. However, the relation between TEs content and mechanical properties of the W is rarely analyzed.

Hydrogen(H) and helium(He) are important elements in the fusion devices. The H/He can lead to the blister formation\cite{Fan2017Current,Shen2017Effects,Zhou2018Helium} and the subsequent degradation of mechanical properties of the W\cite{Zhu2017Studies,Becquart2011Modelling}. This phenomenon is commonly referred to as the H/He embrittlement. There are two main sources of the H/He in the fusion devices: (1) the plasma-background used as the fuel; (2) the transmutation reactions ($n$, $p$) and ($n$, $\alpha$) of the structural material under the fusion neutron irradiation. Recently, the research on the behavior of the H/He in the W has been emphasized and investigated in different perspectives. The formation mechanism of the H/He blisters in the W, the brittle mechanism of the W caused by the H/He, and the diffusion behavior of the H/He in the W have been explored clearly\cite{Lu2014A,Sun2016High,Zhou2015Modeling,Zhou2010Towards,Kong2015First,Wang2015Effects,Becquart2006Migration,Becquart2007Ab,Becquart2009A}. The interaction between 16 kinds of elements (including the transmutation elements) and H/He in the W has been analyzed\cite{Wu2013Effects,Wu2014First,Kong2016First}. The results show that the interaction between the H/He and TEs in the W strongly depends on the charge density perturbation in the vicinity of the solute atom.

In most studies on TEs in the W-based PFM, the effects of Re, Os, and Ta on the crystal structure, mechanical properties and thermodynamic properties of the W have been analyzed. However, a systematic study on the relation of the TEs content and mechanical properties of the W is still lacking. In this paper, the relation between the TEs content and mechanical properties of the W, TEs effects on the formation and diffusion of the H/He defect in the W, and the interaction between the TEs and H/He defect in the W were investigated by using the first-principles calculation method.

\section{Methodologies}
The W belongs to the body-centered cubic (bcc) crystal system, and its space group is Im-3m. Simulation supercell composed of 54 lattice points ($3\times3\times3$) was used in this paper. Lassner\cite{Lassner1999Tungsten} and Jaffee\cite{Jaffee1958The} demonstrated that there is a solid solubility limit of Re and Os in the W-Re, W-Os systems (approximately 27 wt\% Re, 18.5 wt\% Os), respectively. A high atomic percent beyond the above mentioned proportion was chosen to investigate effects of TEs in the W clearly. When TEs effects on mechanical properties of the W were explored, some W atoms in the supercell were replaced by the Re, Os, and Ta atoms in a symmetrical way, and these alloy systems were formulated as W$_{1-x}$Re$_x$, W$_{1-x}$Os$_x$, and W$_{1-x}$Ta$_x$, respectively. The $x$ stands for the atomic percent of TEs in the supercell, the $x$ value varies from 0 to 0.5, inclusively. The bcc structure were maintained in all W-based alloy models in line with the previous theoretical calculations\cite{Ekman2000Phase,Bercegeay2008Second,Muzyk2013First}. The center W atom in the supercell was replaced by a TEs atom (W(Re), W(Os), and W(Ta)). To investigate TEs effects on the behavior of the H/He in the W, the H/He atom was set in the substitutional site (SS), tetrahedral interstitial site (TIS), and octahedral interstitial site (OIS) near the TEs atoms in the W.

Computations in the work were performed by the density functional theory (DFT) and the plane-wave pseudo-potential method\cite{Hohenberg1964Inhomogeneous,Kohn1965Self}, as implemented in the Cambridge sequential total energy package (CASTEP)\cite{Clark2005First}. Generalized gradient approximation (GGA)\cite{Perdew1986Accurate,Perdew1996Generalized} developed by Perdew and Wang (PW91) functional\cite{Burke1998Derivation} was used for describing the exchange-correlation interaction among electrons, and the ultrasoft pseudo-potential were employed for the ion-electron interaction. Since the H atom only has a single valence electron and He atom is a closed-shell atom, the van der Waals force significantly influences TEs effects on the behavior of the H/He atom in the W. Ortmann-Bechstedt-Schmidt(OBS)\cite{Ortmann2006Semiempirical} dispersion correction DFT (DFT-D) were used to describe TEs effects on the behavior of the H/He in W accurately. After the convergence test was completed, basic parameters of the calculation were chosen as follows: The energy cutoff was set to 390 eV for all calculations; space representation equals reciprocal; SCF tolerance equals $1.0\times 10^{-6}$ eV/atom; and $k$ sampling with $5\times5\times5$ $k$-point mesh in the Brillouin zone was used. The atomic positions were determined until these conditions were satisfied: (1) the maximum force on them was lower than 0.05 eV/nm; (2) the maximum change of energy per atom was lower than $1.0\times10^{-5}$ eV; (3) the maximum displacement was lower than 0.001 \AA; and (4) the maximum stress of the crystal was lower than 0.02 GPa.

Re, Os, and Ta elements in the \rm{W}$_{1-x}$\rm{Re}$_x$, \rm{W}$_{1-x}$\rm{Os}$_x$, and \rm{W}$_{1-x}$\rm{Ta}$_x$ alloy models have two possible situations, namely the substitutional site (SS) and interstitial site (IS). Formation energy describes the difficulty of replacing a W atom in the supercell by a TEs atom or insert a TEs atom in the supercell. The formation energy of the TEs in the W is obtained as:
\begin{equation}
E_{\rm{f}}(\rm{W}_{m}\rm{X_n})=\emph{E}_t(\rm{W}_m\rm{X_n})-n\emph{E}(\rm{X})-m\emph{E}(\rm{W})
\end{equation}

where X stands for the TEs, subscripts m and n stand for the number of the W and X atoms in the models. $E\rm{_t(W_mX_n)}$ represents the total energy of the $\rm{W_mX_n}$ supercell; $E(\rm{X})$ stands for the energy of an X atom in bulk X, and $E(\rm{W})$ represents the energy of a W atom in the perfect W crystal, respectively.

Formation energy of a H/He atom in the pure W and $\rm{W_mX_n}$ supercell and binding energy between the TEs and H/He in the W supercell are calculated as follows:

\begin{equation}
E_{\rm{f}}(\rm{W_mA_s})=\emph{E}_{\rm{t}}(\rm{W_mA_s})-m\emph{E}(\rm{W})-\emph{E}(\rm{A})
\end{equation}

\begin{equation}
E_{\rm{f}}(\rm{W_mX_nA_s})=\emph{E}_t(\rm{W_mX_nA_s})-\emph{E}(\rm{A})-\emph{E}(\rm{W_mX_n})
\end{equation}

\begin{equation}
E_{\rm{b}}(\rm{X},\rm{A})=\emph{E}(\rm{W_mX_n})+\emph{E}(\rm{W_yA_s})-\emph{E}(W_mX_nA_s)-y\emph{E}(W)
\end{equation}

Where A stands for the H/He element, subscript s represents the position of the H/He in the supercell, respectively. $E\rm{_t(W_mX_nA_s)}$ stands for the total energy of the $\rm{W_mX_n}$ supercell containing a H/He atom at the s site; $E(\rm{A})$ stands for the energy of an isolated H/He atom; $E(\rm{W_mX_n})$ represents the total energy of $\rm{W_mX_n}$; $E(\rm{W_mA_s})$ is the total energy of the W supercell containing a H/He atom, respectively.

As a bcc metal, mechanical properties of the W are described by three independent elastic constants, namely $C_{11}$, $C_{12}$, and $C_{44}$. When $C_{11}+2C_{12}$\textgreater 0, $C_{44}$\textgreater 0, and $C_{11}-C_{12}$\textgreater 0, the W can exist stably with a bcc structure. The mechanical properties of the W and W-TEs alloys, such as bulk modulus ($B$), shear modulus ($G$), Young's modulus ($E$), and Poisson's ratio ($\nu$), are calculated as follows:
\begin{equation}
B=(C_{11}+2C_{12})/3
\end{equation}

\begin{equation}
G=(C_{11}-C_{12}+3C_{44})/5
\end{equation}

\begin{equation}
E=9BG/(3B+G)
\end{equation}

\begin{equation}
\nu=(3B-2G)/2(3B+G)
\end{equation}

\section{Results and discussion}

To investigate TEs effects on mechanical properties of the W, the formation energy of a Re, Os, and Ta atoms in the W was calculated, the most possible site for the TEs in the W was determined according to calculated results. Tab. 1 shows the formation energy of TEs atom in the W. The results indicate that the formation energy of the Re, Os, and Ta atoms located at the IS site (10.412 eV, 10.103eV, and 11.978 eV) is much greater than that at the SS site (-0.526 eV, -0.016 eV, and -1.04 eV) in the W. These values mean that TEs atoms prefer to stay at the SS site rather than the IS site in the W, the TEs atoms at the SS site are much more stable than that at the IS site in the W in terms of energetics. The results are in agreement with the previous theoretical calculation\cite{Wu2013Effects}. Therefore, TEs atoms were set at the SS site of the W in all the models in this paper.

\begin{figure}[!ht]%[tpb]
\begin{center}
\includegraphics[width=12.0cm,angle=0]{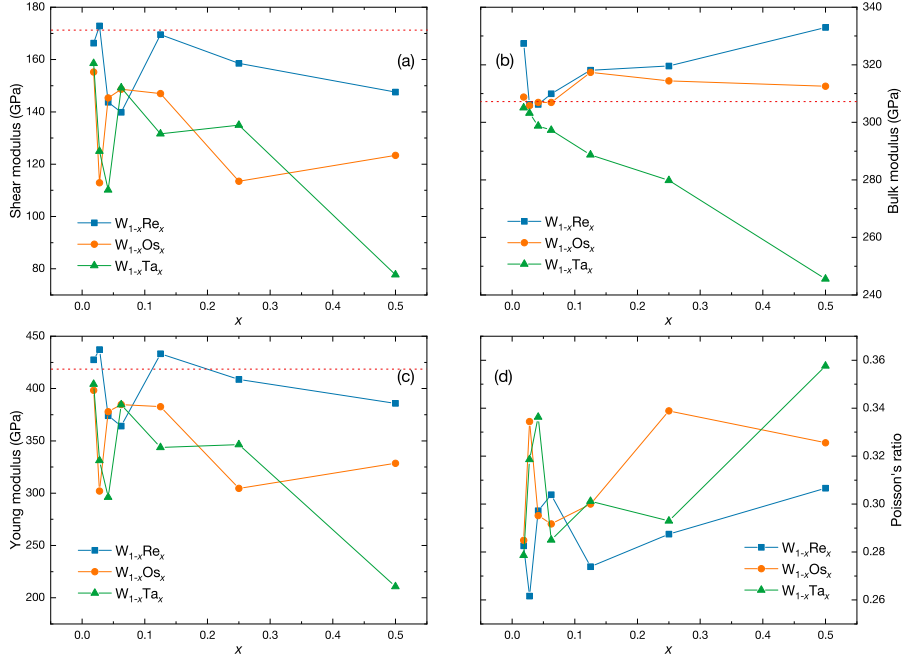}
\caption{Mechanical properties of W$_{1-x}$Re$_x$, W$_{1-x}$Os$_x$, and W$_{1-x}$Ta$_x$, x is the atomic percent of TEs in the W-TEs systems. Panels a, b, c, and d show the shear modulus, bulk modulus, Young's modulus and Poisson's ratio of the W-TEs systems, respectively. The dotted lines stand for the corresponding value of the perfect W.}
\label{Fig.1}
\end{center}
\end{figure}

\begin{table}
\centering
\caption{The formation energy (in eV) of a Re, Os, and Ta atoms in the W.}
\label{table1}
\begin{tabular}{cccc}
\hline
Elements   & Re        & Os      & Ta  \\
\hline
SS         & -0.526    &-0.016   &-1.04\\
IS         & 10.412    &10.103   &11.978\\
\hline
\end{tabular}
\end{table}

\begin{table}
\centering
\caption{Shear modulus ($G$), bulk modulus ($B$), Young's modulus ($E$), and Poisson's ratio ($\nu$) of the perfect W.}
\label{table2}
\begin{tabular}{ccccc}
\hline
Parameters & $G$(GPa)   & $B$(GPa)     & $E$(GPa)   &$\nu$  \\
\hline
W &171.16 &307.23   &418.63 &0.27 \\
Cal.\cite{Becquart2007Ab}    &149.42 &313.61   &386.83 &0.29 \\
Exp.\cite{S1993Theory}       &163.40 &314.33   &417.80 &0.28 \\
\hline
\end{tabular}
\end{table}

As the greatest potential PFM, mechanical properties of the W are significantly important for the stable operation of fusion devices. Tab. 2 shows mechanical properties of the perfect W, the results show that our calculations are in good agreement with the other theoretical\cite{Becquart2007Ab} and experimental\cite{S1993Theory} results. Fig. 1 shows that mechanical properties of W-TEs change with TEs content. The generation of TEs results in the lower shear modulus of the W than that of the perfect W. Re and Os increase the bulk modulus of the W, while the bulk modulus of the W decreases with the Ta concentration. The effects of Re and Os on the bulk modulus of the W are not significant compared with that of Ta atoms. The generation of TEs reduces the Young's modulus of the W, while increases the Poisson's ratio of the W. Mechanical properties of the W change with TEs content. Hence, it is concluded that mechanical properties of the W incompletely depend on TEs content.

\begin{figure}[!ht]%[tpb]
\begin{center}
\includegraphics[width=8.8cm,angle=0]{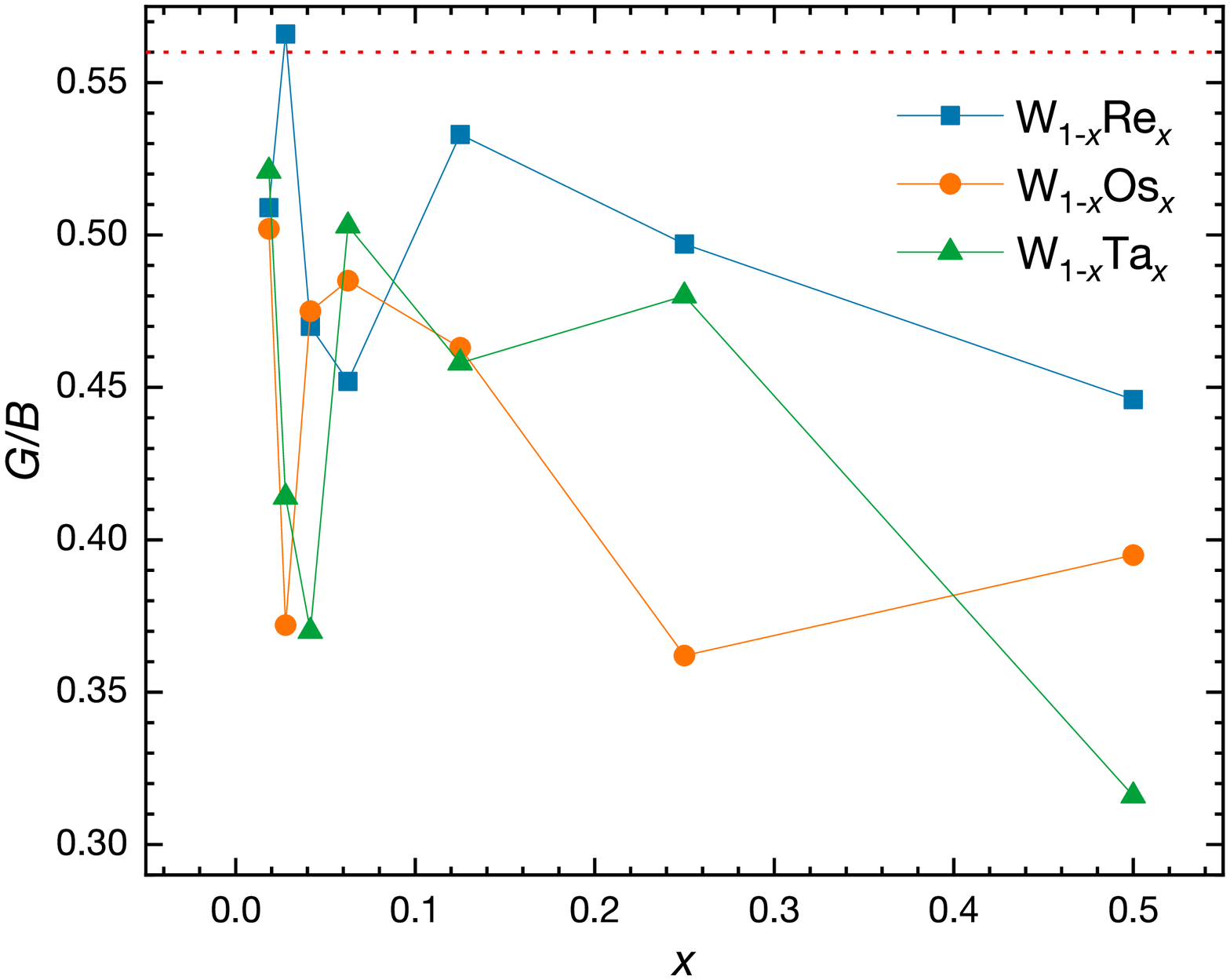}
\caption{The $G/B$ ratio of W$_{1-x}$Re$_x$, W$_{1-x}$Os$_x$, and W$_{1-x}$Ta$_x$. The dotted line stands for the ratio of perfect W.}
\label{Fig.2}
\end{center}
\end{figure}

Pugh presents an empirical relation, which can predict the brittleness and ductility of the bcc metal\cite{S1954XCII}. The empirical formula is the ratio of shear modulus and bulk modulus. The metal is ductile when the ratio is lower than 0.57, and the metal is brittle when the ratio is higher than 0.57. Fig. 2 shows that the ratio of W-TEs varies with TEs content. The ratio of the W is reduced by TEs, indicating that the ductility of the W is enhanced by TEs. The conclusion that TEs can enhance the ductility of the W, is in good agreement with the previous researches\cite{Geach1955pro,Jaffee1958The,Jiang2016The,Li2016Ab,Wei2014First}. It should be noted that the dislocation and other defects were ignored in calculated results in this paper. Even if the change of mechanical properties is the only consideration, the ductility cannot change monotonically with the content of the TEs.

\begin{table}
\centering
\caption{The representative bond population and bond length (in \AA) of two adjacent atoms in the perfect W, W(Re), W(Os), and W(Ta) with some approximate processing.}
\label{table3}
\begin{tabular}{cccc}
\hline
System & Bond   & Length    & Population  \\
\hline
W      & W-W    &2.755      &0.38 \\
W(Re)  &W-W$_1$ &2.751      &0.39 \\
       &W-W$_2$ &2.746      &0.40 \\
       &W-W$_3$ &2.762      &0.39 \\
       &W-Re    &2.751      &0.40 \\
W(Os)  &W-W$_1$ &2.746      &0.39 \\
       &W-W$_2$ &2.746      &0.40 \\
       &W-W$_3$ &2.747      &0.39 \\
       &W-W$_4$ &2.752      &0.39 \\
       &W-W$_5$ &2.754      &0.39 \\
       &W-W$_6$ &2.754      &0.40 \\
       &W-W$_7$ &2.757      &0.39 \\
       &W-W$_8$ &2.759      &0.40 \\
       &W-W$_9$ &2.764      &0.39 \\
       &W-Os    &2.762      &0.36 \\
W(Ta)  &W-W$_1$ &2.729      &0.38 \\
       &W-W$_2$ &2.751      &0.39 \\
       &W-W$_3$ &2.753      &0.39 \\
       &W-W$_4$ &2.753      &0.40 \\
       &W-W$_5$ &2.759      &0.41 \\
       &W-W$_6$ &2.760      &0.42 \\
       &W-W$_7$ &2.761      &0.40 \\
       &W-W$_8$ &2.773      &0.37 \\
       &W-Ta    &2.789      &0.60\\
\hline
\end{tabular}
\end{table}

After TEs are generated in the W, mechanical properties of the W have been changed. To explain the change in mechanical properties of the W, the interaction between different atoms in the W was researched by the bond population. Tab. 3 shows the representative bond population in different models. Compared with perfect W, the generation of TEs, W(Re), W(Os) and W(Ta) result in the greater W-W bond population, and certain shorter distances between two adjacent W atoms. Ta is the left neighbor of W in the periodic table, while Re and Os are on the right side. Hence, alloying with Ta depletes the d-band, while Re and Os fill it. The change in the bond population is one of the factors that can result in the change about mechanical properties of the W.

\begin{table}
\centering
\caption{The formation energy (in eV) of the H/He atom in the perfect W with different configurations.}
\label{table4}
\begin{tabular}{ccccccc}
\hline
Reference &$\rm{H_{SS}}$ & $\rm{H_{TIS}}$ & $\rm{H_{OIS}}$   & $\rm{He_{SS}}$ & $\rm{He_{TIS}}$ & $\rm{He_{OIS}}$  \\
\hline
This paper                   & 1.468 & -2.202 & -1.836 & 4.660 & 5.695 & 5.837  \\
Cal.\cite{Lee2009Energetics}    & 0.92  & -2.47  & -2.07  & 5.00  & 6.23  & 6.48   \\
Cal.\cite{Becquart2007Ab,Skinner2008Recent}       & 0.78  & 2.44   & 2.06   & 4.70  & 6.16  & 6.38   \\
\hline
\end{tabular}
\end{table}

When fusion devices are running, the H/He inevitably enter into the W. The research on TEs effects on the behavior of the H/He in the W is significantly important for the development of the W-based PFM. The formation energy of the H/He atom in the perfect W is shown in Tab. 4 and Fig. 3. The calculation results of this study agree well with the previous theoretical results\cite{Becquart2007Ab,Lee2009Energetics,Skinner2008Recent}. There are some discrepancies between our results and that of other researchers due to different computation methods and the size of the supercell, since the VASP code with a 128-atom supercell was applied in previous works\cite{Kresse1996Efficient,G1996Efficiency}. Although there is a discrepancy in the absolute value of the formation energy, the difference in the formation energy of the H/He atoms with the different sites, such as SS, TIS, and OIS, are in good agreement with each other. The formation energy of a He atom at the SS site in the W has the lowest value (4.66 eV); the H atom has the lowest formation energy (-2.202 eV) when H atom at the TIS site in the W. As a result, the single He atom favors the SS site, while the H atom spontaneously incorporates at the TIS site in W with the most negative formation energy.

\begin{figure}[!ht]%[tpb]
\begin{center}
\includegraphics[width=8.8cm,angle=0]{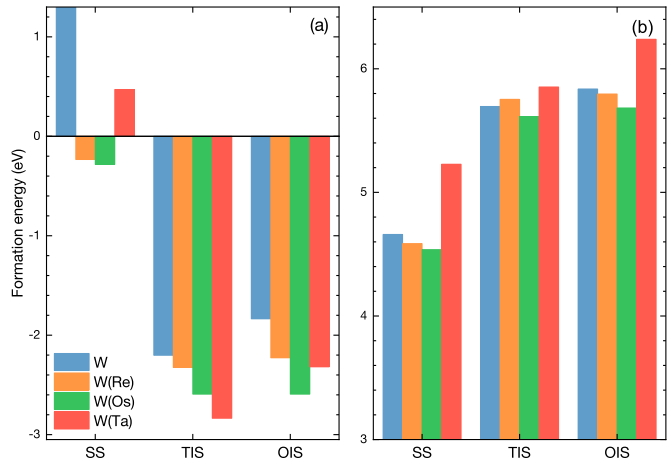}
\caption{The formation energy of the H atom (a) and He atom (b) in the supercell with and without TEs atom.}
\label{Fig.3}
\end{center}
\end{figure}

There is a single electron in the 1$s$ orbit of the H atom, while the He atom has two electrons in the 1$s$ orbit. The H atom has the potential to obtain electrons from its surroundings. The He atom prefers to occupy the low electron density region. The number of the 1$s$ orbital electrons is determined by the Mulliken method\cite{Segall1996Population}. Fig. 4 shows the number of electron in the 1$s$ orbit of the H/He atom. According to our calculations, the closer to 2 the number of electron in the 1$s$ orbit is, the more stable the configuration of the H/He in the W is. When the H atom is located in the TIS site, the number of the electron in the 1$s$ orbit is closer to 2 than other two cases. When the He atom is located at the SS site in W, the number of the electron in the 1$s$ orbit is closer to 2 over other two sites. Therefore, the He atom favors the SS site in the W, while the H atom spontaneously incorporates at the TIS site in the W.

\begin{figure}[!ht]%[tpb]
\begin{center}
\includegraphics[width=8.8cm,angle=0]{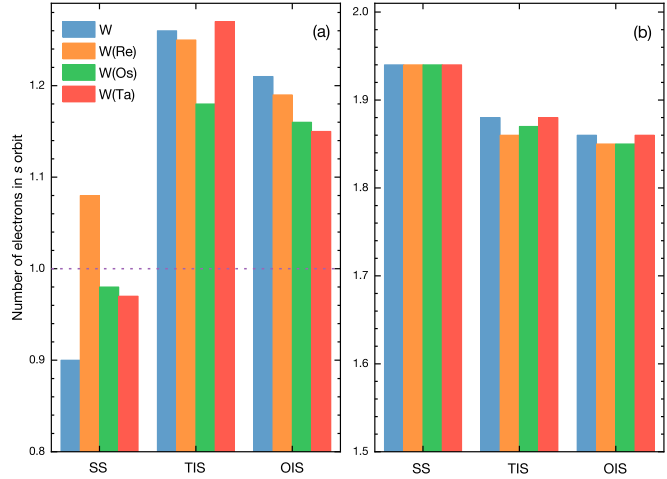}
\caption{ The electron number on the H atom (a) and He (b) atom 1$s$ orbit.}
\label{Fig.4}
\end{center}
\end{figure}

The formation energy of a H/He atom near the TEs atom in the W is shown in Fig. 3. Compared with the formation energy of the H/He atom in the perfect W, the formation energy of the H atom in the W is reduced by TEs, and similar results of theoretical research have been obtained\cite{Wan2018Energetics,Wan2018Hydrogen}. The result suggests that the formation of the substitutional and interstitial H defects near the TEs atom in the W-TEs systems are easier than that in the perfect W, but the H atom still favors the TIS site in the W with the most negative formation energy. Re and Os reduce the formation energy of the substitutional and interstitial He defects near the Re/Os atom in the W, but the Ta increases the formation energy of the He defects in the W. The result indicates that the formation of the He defects in the W is easier due to the generation of Re and Os, while the Ta increases the difficulty of forming the He defects in the W. Although the formation energy of the H/He defect in the W is changed by the generation of TEs, the favorable site of the H/He defect in the W is not changed. The result is in agreement with the conclusion obtained by Liu et. al\cite{Wu2013Effects}. When located at the TIS and SS sites in the W, the number of the 1$s$ electron of the H/He atom is still closer to 2, as shown in Fig. 4.

\begin{figure}[!ht]%[tpb]
\begin{center}
\includegraphics[width=8.8cm,angle=0]{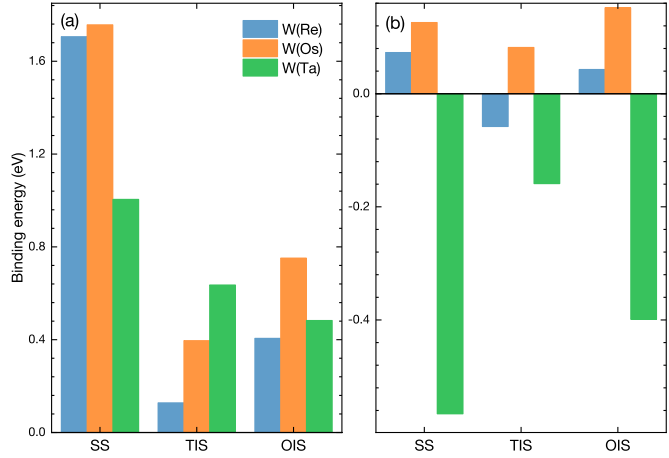}
\caption{The binding energy between the H/He defect and TEs in the W. (a) the binding energy between the H defect and TEs; (b) the binding energy between the He defect and TEs.}
\label{Fig.5}
\end{center}
\end{figure}

\begin{table}
\centering
\caption{The diffusion barrier (in eV) of the H/He in the W with and without the TEs.}
\label{table5}
\begin{tabular}{ccccc}
\hline
~                &Paths      &Path1             &Path2               &path1$^*$  \\
\hline
\multirow{2}*{H} &W          &0.026(0.06$^a$)   &0.144(0.23$^a$)     &--        \\
~                &W(Ta)      &-3.120            &-2.859              &0.061   \\
~                &W(Re)      &2.301             &0.039               &0.027  \\
~                &W(Os)      &0.079             &0.064               &0.015 \\
\multirow{2}*{He}&W          &0.192(0.20$^b$)   &0.346(0.39$^b$)     &--          \\
~                &W(Ta)      &0.203             &0.517               &0.178   \\
~                &W(Re)      &2.580             &0.094               &0.174  \\
~                &W(Os)      &0.286             &0.013               &0.145 \\
\hline
$^a$reference\cite{LI2009FIRST}\\
$^b$ reference\cite{Li2011Stress}\\
\end{tabular}
\end{table}

The positive binding energy indicates that an attractive interaction exists between TEs and H/He in the W, while the negative binding energy indicates a repulsive interaction. The binding energy between TEs and H/He in the W is shown in Fig. 5. The results show that there is an attractive interaction between the TEs and H in the W with a positive binding energy. The attractive interaction between TEs and H at the SS site is stronger than that at the TIS and OIS sites, and the attractive interaction between TEs and H at the TIS site is the weakest. The positive binding energy demonstrates that an attractive interaction exists between Re/Os and He in W, the attractive interaction between the Os and He is stronger than that between the Re and He in the W. It is worth noting that there is a repulsive interaction between the Ta and He in the W with a negative binding energy, the strength of the repulsive interaction between the Ta and He observes the order as follows: SS \textgreater OIS \textgreater TIS. The repulsive interaction decreases with the distance between the Ta and He in the W. The change of binding energy value is in good agreement with the formation energy of the H/He in the W. The attractive interaction between TEs and H is mainly caused by the change of the electron density and the local stress caused by TEs in the supercell. There is a single electron in the H 1$s$ orbit, and the H tends to obtain a charge from its surrounding atoms to keep its stability. According to the previous theoretical research\cite{Wu2014First}, the interaction between TEs and He may be caused by the local stress produced by TEs atoms. The He atom is a closed-shell atom, the charge effects is not significant compared with that of the local stress. The atomic radius of Ta atom is bigger than that of W atom, and the atomic radius of Re and Os atom is smaller than that of W atom. As a result, force fields are generated from the Ta atom to its adjacent W atom, and from the surrounding W atom to the Re/Os atom, respectively.

\begin{figure}[!ht]%[tpb]
\begin{center}
\includegraphics[width=8.8cm,angle=0]{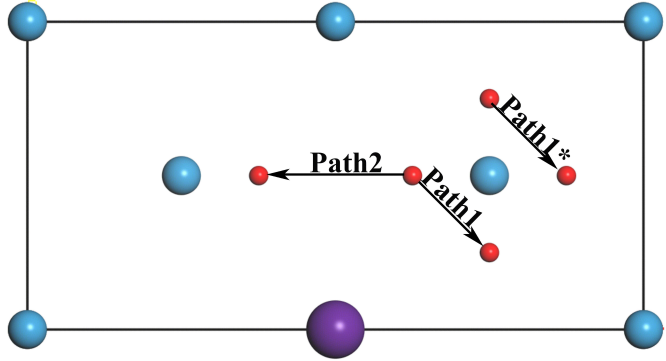}
\caption{The possible diffusion paths of the H/He in the W and W-TEs systems. The blue, purple, and red balls stand for W, TEs, and H/He atoms, respectively.}
\label{Fig.6}
\end{center}
\end{figure}

The most possible diffusion path of the H/He in the W is determined by the diffusion barrier. Each possible diffusion path has a unique diffusion barrier. The possible diffusion paths of the H/He in the W are shown in Fig. 6, Path1 is the H/He diffusion from a TIS to the nearest TIS site (TIS $\rightarrow$ TIS), Path2 is the H/He diffusion from a TIS site to other TIS site through an OIS site (TIS $\rightarrow$ OIS $\rightarrow$ TIS), Path1$^*$ is the same as the Path1 but the Path1$^*$ is a little further from the TEs atom. Note that the diffusion barrier $E_a$ was obtained at 0 K, the diffusion coefficient at different temperatures can be estimated by the Arrhenius diffusion equation $D = D_0 exp(-E_a/k_BT)$, where $D_0$ is the pre-exponential factor and $k_B$ is the Boltzmann constant. According to the equation, a higher diffusion barrier indicates a lower diffusion rate. The diffusion barrier of the H/He in the W with and without the TEs was calculated, respectively. Tab. 5 shows the diffusion barrier of the H/He in the W through the different diffusion paths, calculated results are in agreement with the previous DFT research\cite{Li2011Stress,LI2009FIRST}. In the pure W, the diffusion barrier of the H/He through the Path1 is lower than that of the Path2, the result indicates that the H/He prefers to diffuse through the Path1 in the W. The diffusion barrier of the He is higher than that of the H in the W, indicating that diffusion rate of the He is lower than that of the H in the W. As shown in the Tab. 5, the Re/Os increases the diffusion barrier of the H through the Path1, while decreases the diffusion barrier through the Path2. The diffusion barrier of the H near the Ta through the path1 and Path2 is negative, indicating that there is no barrier in these two paths for the diffusion of H. The diffusion barrier of the H through Path1$^*$ is lower than that of Path1 and Path2, indicating that the diffusion rate of the H in the W increases with the distance between the H and TEs. The diffusion barrier of He through the Path1 is increased by TEs, the diffusion rate of He in the W-TEs through the Path1 is lower than that in the perfect W. The Ta increases the diffusion barrier of He through the Path2, while the Re/Os lower the diffusion barrier. The best diffusion path of the He in W is changed into the Path1$^*$ by the Ta, because of the repulsive interaction between He and Ta. The best diffusion path of the He is changed into the Path2 in W by Re/Os. This is caused by the attractive interaction between He and Re/Os. All of the results show that the interaction between TEs and H/He in the W plays an important role in the change of the best diffusion path. In general, the diffusion rate of the H/He can be increased by TEs in the W.

\section{Conclusion}

In this paper, TEs (Re, Os, and Ta) effects on mechanical properties of the W and behavior of the H/He in the W were investigated by using the first-principles calculation method. The results show that TEs can enhance the ductility of W due to the increased bond population and the decreased distance between two adjacent W atoms. Besides, there is no monotonous linear relationship between mechanical properties of W-TEs system and TEs content. The formation energy of the H/He defects in the W is decreased by the Re/Os. The Ta can decrease the formation energy of the H defect but increase the formation energy of the He defect in the W. The difficulty of forming a He defect in the W is increased by the Ta, while the difficulty  of forming the H/He defect in the W is reduced by TEs. The binding energy indicates that there is an attractive interaction between TEs and H/He in the W except for the repulsive interaction between Ta and He in W. TEs can change the best diffusion path of H/He in W due to the interaction between the H/He and TEs. The diffusion rate increases with the distance between the H/He and TEs. All the results show that TEs effects should be further discussed in the research of W-based PFM.

\section*{Data availability}

All data included in this study are available upon request by contact with the corresponding author.

\section*{Acknowledgments}

This work was supported by the Fundamental Research Funds for the Central Universities under Grant Nos. 2017MS079 and 2018ZD10, and the National Natural Science Foundation of China under Grant No. 11705059.

\section*{References}



  \bibliographystyle{elsarticle-num}
 \bibliography{mybib}

\begin{thebibliography}{10}
\expandafter\ifx\csname url\endcsname\relax
  \def\url#1{\texttt{#1}}\fi
\expandafter\ifx\csname urlprefix\endcsname\relax\def\urlprefix{URL }\fi
\expandafter\ifx\csname href\endcsname\relax
  \def\href#1#2{#2} \def\path#1{#1}\fi

\bibitem{Lawson1957Power}
J.~D. Lawson, Nat. 180~(4590) (1957) 780--782.
\newblock \href{https://doi.org/10.1038/180780a0}{[link]}.
\newline\urlprefix\url{https://doi.org/10.1038/180780a0}

\bibitem{Philipps2011Tungsten}
V.~Philipps, J. Nucl. Mater. 415~(1) (2011) S2–S9.
\newblock \href{https://doi.org/10.1016/j.jnucmat.2011.01.110}{[link]}.
\newline\urlprefix\url{https://doi.org/10.1016/j.jnucmat.2011.01.110}

\bibitem{Cottrell2006Transmutation}
G.~A. Cottrell, R.~Pampin, N.~P. Taylor, G.~A. Cottrell, R.~Pampin, N.~P.
  Taylor, Fusion Sci. Technol. 50~(1) (2006) 89--98.
\newblock \href{https://doi.org/10.13182/fst06-a1224}{[link]}.
\newline\urlprefix\url{https://doi.org/10.13182/fst06-a1224}

\bibitem{Gilbert2011Neutron}
M.~R. Gilbert, J.~C. Sublet, Nucl. Fusion 51~(4) (2011) 043005.
\newblock \href{https://doi.org/10.1088/0029-5515/51/4/043005}{[link]}.
\newline\urlprefix\url{https://doi.org/10.1088/0029-5515/51/4/043005}

\bibitem{Hasegawa2014Neutron}
A.~Hasegawa, M.~Fukuda, S.~Nogami, K.~Yabuuchi, Nucl. Eng. Des. 89~(7–8)
  (2014) 1568--1572.
\newblock \href{https://doi.org/10.1016/j.fusengdes.2014.04.035}{[link]}.
\newline\urlprefix\url{https://doi.org/10.1016/j.fusengdes.2014.04.035}

\bibitem{Hasegawa2015Neutron}
A.~Hasegawa, M.~Fukuda, K.~Yabuuchi, S.~Nogami, J. Nucl. Mater. 471 (2015)
  175--183.
\newblock \href{https://doi.org/10.1016/j.jnucmat.2015.10.047}{[link]}.
\newline\urlprefix\url{https://doi.org/10.1016/j.jnucmat.2015.10.047}

\bibitem{freysoldt2014first}
C.~Freysoldt, B.~Grabowski, T.~Hickel, J.~Neugebauer, G.~Kresse, A.~Janotti,
  C.~G. Van~de Walle, Reviews of modern physics 86~(1) (2014) 253.
\newblock \href{https://doi.org/10.1103/RevModPhys.86.253}{[link]}.
\newline\urlprefix\url{https://doi.org/10.1103/RevModPhys.86.253}

\bibitem{wimmer1995computational}
E.~Wimmer, Science 269~(5229) (1995) 1397--1399.
\newblock \href{https://doi.org/10.1126/science.269.5229.1397}{[link]}.
\newline\urlprefix\url{https://doi.org/10.1126/science.269.5229.1397}

\bibitem{li2018effects}
H.~Li, K.~Shin, G.~Henkelman, The Journal of chemical physics 149~(17) (2018)
  174705.
\newblock \href{https://doi.org/10.1063/1.5053894}{[link]}.
\newline\urlprefix\url{https://doi.org/10.1063/1.5053894}

\bibitem{li2017dehydrogenation}
H.~Li, G.~Henkelman, The Journal of Physical Chemistry C 121~(49) (2017)
  27504--27510.
\newblock \href{https://doi.org/10.1021/acs.jpcc.7b09953}{[link]}.
\newline\urlprefix\url{https://doi.org/10.1021/acs.jpcc.7b09953}

\bibitem{chouhan2012ab}
R.~K. Chouhan, A.~Alam, S.~Ghosh, A.~Mookerjee, Journal of Physics: Condensed
  Matter 24~(37) (2012) 375401.
\newblock \href{https://doi.org/10.1088/0953-8984/24/37/375401}{[link]}.
\newline\urlprefix\url{https://doi.org/10.1088/0953-8984/24/37/375401}

\bibitem{Giusepponi2013The}
S.~Giusepponi, M.~Celino, J. Nucl. Mater. 435~(1–3) (2013) 52--55.
\newblock \href{https://doi.org/10.1016/j.jnucmat.2012.12.028}{[link]}.
\newline\urlprefix\url{https://doi.org/10.1016/j.jnucmat.2012.12.028}

\bibitem{Wei2014First}
N.~Wei, T.~Jia, X.~Zhang, T.~Liu, Z.~Zeng, X.~Y. Yang, AIP Adv. 4~(5) (2014)
  057103.
\newblock \href{https://doi.org/10.1063/1.4875024}{[link]}.
\newline\urlprefix\url{https://doi.org/10.1063/1.4875024}

\bibitem{Li2016Ab}
X.~Li, S.~Sch\"onecker, R.~Li, X.~Li, Y.~Wang, J.~Zhao, B.~Johansson, L.~Vitos,
  Journal of Physics: Condensed Matter 28~(29) (2016) 295501.
\newblock \href{https://doi.org/10.1088/0953-8984/28/29/295501}{[link]}.
\newline\urlprefix\url{https://doi.org/10.1088/0953-8984/28/29/295501}

\bibitem{Bonny2017Elastic}
G.~Bonny, A.~Bakaev, D.~Terentyev, Y.~A. Mastrikov, Scripta Mater. 128 (2017)
  45--48.
\newblock \href{https://doi.org/10.1016/j.scriptamat.2016.09.039}{[link]}.
\newline\urlprefix\url{https://doi.org/10.1016/j.scriptamat.2016.09.039}

\bibitem{Setyawan2016Ab}
W.~Setyawan, G.~Nandipati, R.~J. Kurtz, J. Nucl. Mater. 484 (2017) 30--41.
\newblock \href{https://doi.org/10.1016/j.jnucmat.2016.11.002}{[link]}.
\newline\urlprefix\url{https://doi.org/10.1016/j.jnucmat.2016.11.002}

\bibitem{Hossain2014Stress}
M.~Z. Hossain, J.~Marian, Acta Mater. 80 (2014) 107--117.
\newblock \href{https://doi.org/10.1016/j.actamat.2014.07.028}{[link]}.
\newline\urlprefix\url{https://doi.org/10.1016/j.actamat.2014.07.028}

\bibitem{wrobel2017first}
J.~Wr{\'o}bel, D.~Nguyen-Manh, K.~Kurzyd{\l}owski, S.~Dudarev,
  \href{https://doi.org/10.1088/1361-648X/aa5f37}{A first-principles model for
  anomalous segregation in dilute ternary tungsten-rhenium-vacancy alloys},
  Journal of Physics: Condensed Matter 29~(14) (2017) 145403.
\newline\urlprefix\url{https://doi.org/10.1088/1361-648X/aa5f37}

\bibitem{Fan2017Current}
H.~Fan, T.~Endo, Z.~Bi, W.~Yan, S.~Ohnuki, Q.~Yang, W.~Ni, D.~Liu, J. Nucl.
  Mater. 486 (2017) 191--196.
\newblock \href{https://doi.org/10.1016/j.jnucmat.2017.01.025}{[link]}.
\newline\urlprefix\url{https://doi.org/10.1016/j.jnucmat.2017.01.025}

\bibitem{Shen2017Effects}
Z.~Shen, Z.~Zheng, F.~Luo, W.~Hu, W.~Zhang, L.~Guo, Y.~Ren, Fusion Eng. Des.
  115 (2017) 80--84.
\newblock \href{https://doi.org/10.1016/j.fusengdes.2017.01.001}{[link]}.
\newline\urlprefix\url{https://doi.org/10.1016/j.fusengdes.2017.01.001}

\bibitem{Zhou2018Helium}
Q.~Zhou, K.~Azuma, A.~Togari, M.~Yajima, M.~Tokitani, S.~Masuzaki, N.~Yoshida,
  M.~Hara, Y.~Hatano, Y.~Oya, J. Nucl. Mater. 502 (2018) 289--294.
\newblock \href{https://doi.org/10.1016/j.jnucmat.2018.02.035}{[link]}.
\newline\urlprefix\url{https://doi.org/10.1016/j.jnucmat.2018.02.035}

\bibitem{Zhu2017Studies}
K.~Zhu, Y.~Xing, T.~Liu, R.~Yang, Z.~Long, H.~Zhou, L.~Luo, P.~Zhou, X.~Ye,
  W.~Wang, Mater. Lett. 213 (2017) 362--365.
\newblock \href{https://doi.org/10.1016/j.matlet.2017.11.028}{[link]}.
\newline\urlprefix\url{https://doi.org/10.1016/j.matlet.2017.11.028}

\bibitem{Becquart2011Modelling}
C.~S. Becquart, M.~F. Barthe, A.~De~Backer, Phys. Scripta 2011~(T145) (2011)
  1972--1978.
\newblock \href{https://doi.org/10.1088/0031-8949/2011/T145/014048}{[link]}.
\newline\urlprefix\url{https://doi.org/10.1088/0031-8949/2011/T145/014048}

\bibitem{Lu2014A}
G.~H. Lu, H.~B. Zhou, C.~S. Becquart, Nucl. Fusion 54~(54) (2014) 086001.
\newblock \href{https://doi.org/10.1088/0029-5515/54/8/086001}{[link]}.
\newline\urlprefix\url{https://doi.org/10.1088/0029-5515/54/8/086001}

\bibitem{Sun2016High}
L.~Sun, S.~Jin, G.~H. Lu, L.~Wang, Scripta Mater. 122 (2016) 14--17.
\newblock \href{https://doi.org/10.1016/j.scriptamat.2016.05.007}{[link]}.
\newline\urlprefix\url{https://doi.org/10.1016/j.scriptamat.2016.05.007}

\bibitem{Zhou2015Modeling}
H.~B. Zhou, Y.~H. Li, G.~H. Lu, Comp. Mater. Sci. 112 (2015) 487--491.
\newblock \href{https://doi.org/10.1016/j.commatsci.2015.09.019}{[link]}.
\newline\urlprefix\url{https://doi.org/10.1016/j.commatsci.2015.09.019}

\bibitem{Zhou2010Towards}
H.~B. Zhou, Y.~L. Liu, S.~Jin, Y.~Zhang, G.~N. Luo, G.~H. Lu, Nucl. Fusion
  50~(11) (2010) 2981--2989.
\newblock \href{https://doi.org/10.1088/0029-5515/50/11/115010}{[link]}.
\newline\urlprefix\url{https://doi.org/10.1088/0029-5515/50/11/115010}

\bibitem{Kong2015First}
X.~S. Kong, S.~Wang, X.~Wu, Y.~W. You, C.~S. Liu, Q.~F. Fang, J.~L. Chen, G.~N.
  Luo, Acta Mater. 84 (2015) 426--435.
\newblock \href{https://doi.org/10.1016/j.actamat.2014.10.039}{[link]}.
\newline\urlprefix\url{https://doi.org/10.1016/j.actamat.2014.10.039}

\bibitem{Wang2015Effects}
S.~Wang, X.~S. Kong, X.~Wu, Q.~F. Fang, J.~L. Chen, G.~N. Luo, C.~S. Liu, J.
  Nucl. Mater. 459~(6) (2015) 143--149.
\newblock \href{https://doi.org/10.1016/j.jnucmat.2015.01.025}{[link]}.
\newline\urlprefix\url{https://doi.org/10.1016/j.jnucmat.2015.01.025}

\bibitem{Becquart2006Migration}
C.~S. Becquart, Phys. Rev. Lett. 97~(19) (2006) 196402.
\newblock \href{https://doi.org/10.1103/PhysRevLett.97.196402}{[link]}.
\newline\urlprefix\url{https://doi.org/10.1103/PhysRevLett.97.196402}

\bibitem{Becquart2007Ab}
C.~S. Becquart, C.~Domain, Nucl. Instrum. Meth. B 255~(1) (2007) 23--26.
\newblock \href{https://doi.org/10.1016/j.nimb.2006.11.006}{[link]}.
\newline\urlprefix\url{https://doi.org/10.1016/j.nimb.2006.11.006}

\bibitem{Becquart2009A}
C.~S. Becquart, C.~Domain, J. Nucl. Mater. s386–388~(5) (2009) 109--111.
\newblock \href{https://doi.org/10.1016/j.jnucmat.2008.12.085}{[link]}.
\newline\urlprefix\url{https://doi.org/10.1016/j.jnucmat.2008.12.085}

\bibitem{Wu2013Effects}
X.~Wu, X.~S. Kong, Y.~W. You, C.~S. Liu, Q.~F. Fang, J.~L. Chen, G.~N. Luo,
  Z.~Wang, Nucl. Fusion 53~(7) (2013) 073049.
\newblock \href{https://doi.org/10.1088/0029-5515/53/7/073049}{[link]}.
\newline\urlprefix\url{https://doi.org/10.1088/0029-5515/53/7/073049}

\bibitem{Wu2014First}
X.~Wu, X.~S. Kong, Y.~W. You, C.~S. Liu, Q.~F. Fang, J.~L. Chen, G.~N. Luo,
  Z.~Wang, J. Nucl. Mater. 455~(1–3) (2014) 151--156.
\newblock \href{https://doi.org/10.1016/j.jnucmat.2014.05.060}{[link]}.
\newline\urlprefix\url{https://doi.org/10.1016/j.jnucmat.2014.05.060}

\bibitem{Kong2016First}
X.~S. Kong, X.~Wu, C.~S. Liu, Q.~F. Fang, Q.~M. Hu, J.~L. Chen, G.~N. Luo,
  Nucl. Fusion 56~(2) (2016) 026004.
\newblock \href{https://doi.org/10.1088/0029-5515/56/2/026004}{[link]}.
\newline\urlprefix\url{https://doi.org/10.1088/0029-5515/56/2/026004}

\bibitem{Lassner1999Tungsten}
E.~Lassner, W.~D. Schubert,
  \href{https://doi.org/10.1007/978-1-4615-4907-9}{Tungsten: Properties,
  chemistry, technology of the element, alloys, and chemical compounds}, Ph.D.
  thesis, Springer, Berlin (1999).
\newline\urlprefix\url{https://doi.org/10.1007/978-1-4615-4907-9}

\bibitem{Jaffee1958The}
R.~I. Jaffee, C.~T. Sims, J.~Harwood, The effect of rhenium on the
  fabricability and ductility of molybdenum and tungsten, Vol. 3rd Plansee
  Seminar Proceedings, Plansee AG, Reutte, 1958.

\bibitem{Ekman2000Phase}
M.~Ekman, K.~Persson, G.~Grimvall, J. Nucl. Mater. 278~(2) (2000) 273--276.
\newblock \href{https://doi.org/10.1016/S0022-3115(99)00241-X}{[link]}.
\newline\urlprefix\url{https://doi.org/10.1016/S0022-3115(99)00241-X}

\bibitem{Bercegeay2008Second}
C.~Bercegeay, G.~Jomard, S.~Bernard, Phys. Rev. B 77~(10) (2008) 4203.
\newblock \href{https://doi.org/10.1103/PhysRevB.77.104203}{[link]}.
\newline\urlprefix\url{https://doi.org/10.1103/PhysRevB.77.104203}

\bibitem{Muzyk2013First}
M.~Muzyk, D.~Nguyen-Manh, J.~Wróbel, K.~J. Kurzydłowski, N.~L. Baluc, S.~L.
  Dudarev, J. Nucl. Mater. 442~(1–3) (2013) S680--S683.
\newblock \href{https://doi.org/10.1016/j.jnucmat.2012.10.025}{[link]}.
\newline\urlprefix\url{https://doi.org/10.1016/j.jnucmat.2012.10.025}

\bibitem{Hohenberg1964Inhomogeneous}
P.~Hohenberg, W.~Kohn, Phys. Rev. 136~(3) (1964) B864.
\newblock \href{https://doi.org/10.1103/PhysRev.136.B864}{[link]}.
\newline\urlprefix\url{https://doi.org/10.1103/PhysRev.136.B864}

\bibitem{Kohn1965Self}
W.~Kohn, L.~J. Sham, Phys. Rev. 140~(4A) (1965) A1133--A1138.
\newblock \href{https://doi.org/10.1103/PhysRev.140.A1133}{[link]}.
\newline\urlprefix\url{https://doi.org/10.1103/PhysRev.140.A1133}

\bibitem{Clark2005First}
S.~J. Clark, M.~D. Segall, C.~J. Pickard, P.~J. Hasnip, M.~I.~J. Probert,
  K.~Refson, M.~C. Payne, Z. Kristallogr. 220~(5/6) (2005) 567--570.
\newblock \href{https://doi.org/10.1524/zkri.220.5.567.65075}{[link]}.
\newline\urlprefix\url{https://doi.org/10.1524/zkri.220.5.567.65075}

\bibitem{Perdew1986Accurate}
J.~P. Perdew, W.~Yue, Phys. Rev. B 33~(12) (1986) 8800--8802.
\newblock \href{https://doi.org/10.1103/physrevb.33.8800}{[link]}.
\newline\urlprefix\url{https://doi.org/10.1103/physrevb.33.8800}

\bibitem{Perdew1996Generalized}
J.~P. Perdew, K.~Burke, M.~Ernzerhof, Phy. Rev. Lett. 77~(18) (1996)
  3865--3868.
\newblock \href{https://doi.org/10.1103/physrevlett.77.3865}{[link]}.
\newline\urlprefix\url{https://doi.org/10.1103/physrevlett.77.3865}

\bibitem{Burke1998Derivation}
K.~Burke, J.~P. Perdew, Y.~Wang, Springer US, 1998.
\newblock \href{https://doi.org/10.1007/978-1-4899-0316-7_7}{[link]}.
\newline\urlprefix\url{https://doi.org/10.1007/978-1-4899-0316-7_7}

\bibitem{Ortmann2006Semiempirical}
F.~Ortmann, F.~Bechstedt, W.~G. Schmidt, Phys. Rev. B 73~(20) (2006) 205101.
\newblock \href{https://doi.org/10.1103/physrevb.73.205101}{[link]}.
\newline\urlprefix\url{https://doi.org/10.1103/physrevb.73.205101}

\bibitem{S1993Theory}
P.~S\"oderlind, O.~Eriksson, J.~M. Wills, A.~M. Boring, Phys. Rev. B 48~(9)
  (1993) 5844--5851.
\newblock \href{https://doi.org/10.1103/physrevb.48.5844}{[link]}.
\newline\urlprefix\url{https://doi.org/10.1103/physrevb.48.5844}

\bibitem{S1954XCII}
S.~F. Pugh, Philos. Mag. 45~(367) (1954) 823--843.
\newblock \href{https://doi.org/10.1080/14786440808520496}{[link]}.
\newline\urlprefix\url{https://doi.org/10.1080/14786440808520496}

\bibitem{Geach1955pro}
G.~Geachand, J.~Hughes, The alloy of rhenium with molybdenum or with tungsten
  and having good high temperature properties, Vol. Proceedings of the Second
  Plansee Seminar, Plansee AG, Reutte, 1955.

\bibitem{Jiang2016The}
D.~Jiang, Q.~Wang, W.~Hu, Z.~Wei, J.~Tong, H.~Wan, J. Mater. Res. 31~(21)
  (2016) 3401--3408.
\newblock \href{https://doi.org/10.1557/jmr.2016.358}{[link]}.
\newline\urlprefix\url{https://doi.org/10.1557/jmr.2016.358}

\bibitem{Lee2009Energetics}
S.~C. Lee, J.~H. Choi, J.~G. Lee, J. Nucl. Mater. 383~(3) (2009) 244--246.
\newblock \href{https://doi.org/10.1016/j.jnucmat.2008.09.017}{[link]}.
\newline\urlprefix\url{https://doi.org/10.1016/j.jnucmat.2008.09.017}

\bibitem{Skinner2008Recent}
C.~H. Skinner, A.~A. Haasz, V.~K.~H. Almov, N.~Bekris, R.~A. Causey, R.~E.~H.
  Clark, J.~P. Coad, J.~W. Davis, R.~P. Doerner, M.~Mayer, Fusion Sci. Technol.
  54~(4) (2008) 891--945.
\newblock \href{https://doi.org/10.13182/fst08-25}{[link]}.
\newline\urlprefix\url{https://doi.org/10.13182/fst08-25}

\bibitem{Kresse1996Efficient}
G.~Kresse, J.~Furthm\"uller, Phys. Rev. B 54~(16) (1996) 11169--11186.
\newblock \href{https://doi.org/10.1103/PhysRevB.54.11169}{[link]}.
\newline\urlprefix\url{https://doi.org/10.1103/PhysRevB.54.11169}

\bibitem{G1996Efficiency}
G.~Kresse, F.~J., Comp. mater. sci. 6~(1) (1996) 15--50.
\newblock \href{https://doi.org/10.1016/0927-0256(96)00008-0}{[link]}.
\newline\urlprefix\url{https://doi.org/10.1016/0927-0256(96)00008-0}

\bibitem{Segall1996Population}
M.~D. Segall, R.~Shah, C.~J. Pickard, M.~C. Payne, Phys. Rev. B 54~(23) (1996)
  16317.
\newblock \href{https://doi.org/10.1103/PhysRevB.54.16317}{[link]}.
\newline\urlprefix\url{https://doi.org/10.1103/PhysRevB.54.16317}

\bibitem{Wan2018Energetics}
C.~B. Wan, S.~Y. Yu, X.~Ju, Chin. Phys. Lett. 35~(4) (2018) 047102.
\newblock \href{https://doi.org/10.1088/0256-307X/35/4/047102}{[link]}.
\newline\urlprefix\url{https://doi.org/10.1088/0256-307X/35/4/047102}

\bibitem{Wan2018Hydrogen}
C.~B. Wan, S.~Y. Yu, X.~Ju, W.~W. Wang, J. Nucl. Mater. 508 (2018) 249--256.
\newblock \href{https://doi.org/10.1016/j.jnucmat.2018.05.050}{[link]}.
\newline\urlprefix\url{https://doi.org/10.1016/j.jnucmat.2018.05.050}

\bibitem{LI2009FIRST}
L.~Yang, H.~K. Liu, X.~T. Zu, Int. J. Mod. Phys. B 23~(08) (2009) 2077--2082.
\newblock \href{https://doi.org/10.1142/S0217979209049334}{[link]}.
\newline\urlprefix\url{https://doi.org/10.1142/S0217979209049334}

\bibitem{Li2011Stress}
W.~Y. Li, Y.~Zhang, H.~B. Zhou, S.~Jin, G.~H. Lu, Nucl. Instrum. Meth. B
  269~(14) (2011) 1731--1734.
\newblock \href{dx.doi.org/10.1016/j.nimb.2010.12.027}{[link]}.
\newline\urlprefix\url{dx.doi.org/10.1016/j.nimb.2010.12.027}

\end{thebibliography}


%\begin{thebibliography}{00}


%\bibitem{}

%\end{thebibliography}
\end{document}